\documentstyle[preprint,aps,epsfig,floats]{revtex}
\begin{document}
\draft
\title{Coupled Growing Networks}
\author{Dafang Zheng$^{1,2,}$\thanks{e-mail:
Dafang.Zheng@brunel.ac.uk} and G\"{u}ler  Erg\"{u}n$^{1}$}
\address{$^{1}$ Department of Mathematical Sciences, Brunel  University,\\
 Uxbridge, Middlesex UB8 3PH, UK}
\address{$^{2}$ Department of Applied Physics, South China University of
Technology,\\ Guangzhou 510641, P.R. China} \maketitle

\address{Electronic address: masrddz@brunel.ac.uk}
\maketitle
\thispagestyle{empty}

\begin{abstract}

We introduce and solve a model which considers  two coupled
networks growing simultaneously. The dynamics of the networks is
governed by the new arrival of network elements (nodes) making
preferential attachments to pre-existing nodes in both networks.
The model segregates the links in the networks as intra-links,
cross-links and mix-links. The corresponding degree distributions
of these links are found to be power-laws with  exponents having
coupled parameters for intra- and cross-links. In the weak
coupling case the model reduces to a simple citation network. As
for the strong coupling, it mimics the mechanism of \emph{the web
of human sexual contacts}.
\end{abstract}


\pagebreak

\section{Introduction}
\label{sec:intro} Today, with a vast amount of publications being
produced in every discipline of scientific research, it can be
rather overwhelming to select a good quality work; that is
enriched with original ideas and relevant to scientific community.
More often this type of publications are discovered through the
citation mechanism. It is believed that an estimate measure for
scientific credibility of a paper is the number of citations that
it receives, though this should not be taken too literally since
some publications may have gone unnoticed or have been forgotten
about over time.

Knowledge of how many times their publications are cited can be
seen as good feedback for the authors, which brings about an
unspoken demand for the statistical analysis of citation data. One
of the impressive empirical studies on citation distribution of
scientific publications \cite{citation} showed that the
distribution is a power-law form with exponent $\gamma \approx 3$.
The power-law behaviour in this complex system is a consequence of
highly cited papers being more likely to acquire further
citations. This was identified as a \emph{preferential attachment}
process in \cite{barab}.

The citation distribution of scientific publications is well
studied and there exist a number of network models
\cite{connectivity,exactsol,bilke} to mimic its complex structure
and empirical results \cite{citation,vazquez} to confirm
predictions. However, they seem to concentrate on the total number
of citations without giving information about the issuing
publications. The scientific publications belonging to a
particular research area do not restrict their references to that
discipline only, they form bridges by comparing or confirming
findings in other research fields. For instance most \emph{Small
World Network Models} \cite{bestconnected,newman,watts} presented
in statistical mechanics, reference a sociometry article
\cite{milgram} which presents the studies of Milgram on the small
world problem. This is the type of process which we will
investigate with a simple model that only considers two research
areas and referencing within and across each other. The
consideration of cross linking also makes the model applicable to
\emph{the web of human sexual contacts} \cite{liljeros,guler},
where  the interactions between males and females can be thought
of as two coupled growing networks.

This paper is organized as follows: In the proceeding section the
model is defined and analyzed with a rate equation approach
\cite{organization,ergun}. In the final section discussions and
comparisons of findings with the existing data are presented.
\section{The model}
\label{sec:the model}

\begin{figure}
\centerline{\epsfig{file=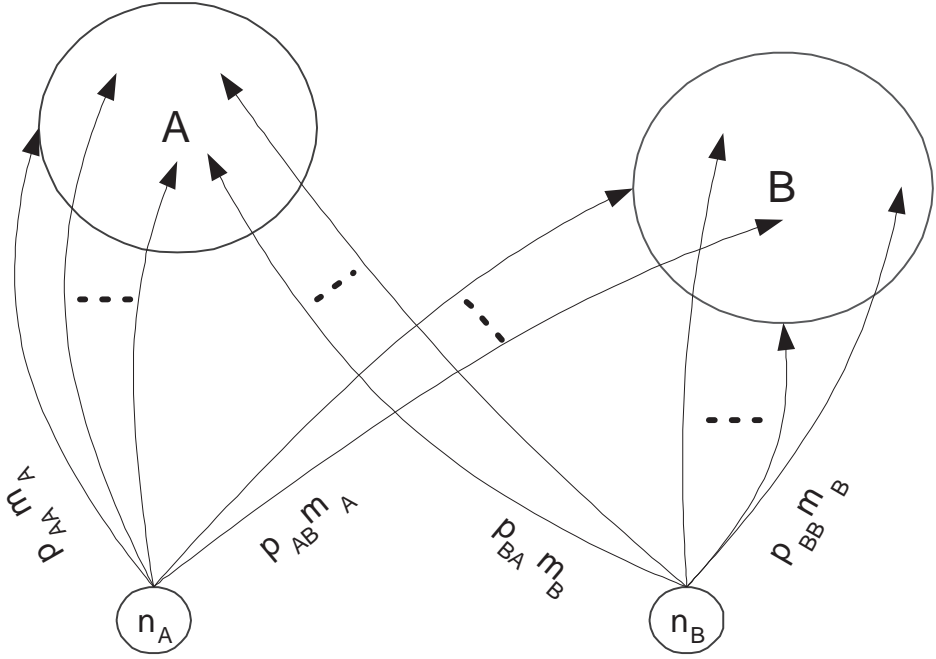,width=14.5cm}}
\vspace{12pt} \caption{\noindent Schematic illustration of the
growth processes of network $A$ and $B$. Each new node $n_{A}$ or
$n_{B}$ is
making a fractional cross linking to other network as
well as intra links to their own .}\label{coupled}
\end{figure}

One can visualize the proposed model with the aid of
Fig.~(\ref{coupled}) that attempts to illustrate the growth
mechanism. We build the model by the following considerations.

Initially, both networks $A$ and $B$ contains $m_{0}$ nodes with
no cross-links between the nodes in the networks. At each time
step two new nodes with no incoming links, one belonging to
network $A$ and the other to $B$, are introduced simultaneously.
The new node joining to $A$ with $m_{A}\leq m_{0}$ outgoing links,
attaches $p_{AA}$ fraction of its links to pre-existing nodes in
$A$ and $p_{AB}=1-p_{AA}$ fraction of them to pre-existing nodes
in $B$. The similar process takes place when a new node joins to
$B$, where the new node has $m_{B}\leq m_{0}$ outgoing links from
which $p_{BB}$ of them goes to nodes in $B$ and the complementary
$p_{BA}=1-p_{BB}$ goes to $A$. The attachments to nodes in either
networks are preferential and the rate of acquiring a link depends
on the number of connections and the initial attractiveness of the
pre-existing nodes.

We define $N_{A}(k_{A},t)$ as the average number of nodes with
total $k_{A}$ number of connections that includes the incoming
intra-links $k_{AA}$ and the incoming cross-links $k_{BA}$  in
network $A$ at time $t$. Similarly, $N_{B}(k_{B},t)$ is the
average number of nodes with $k_{B}=k_{BB}+k_{AB}$ connections at
time $t$ in network $B$. Notice that the indices are
discriminative and the order in which they are used is important,
as they indicate the direction that the links are made. Further
more we also define $N_{AA}(k_{AA},t)$ and $N_{BB}(k_{BB},t)$ the
average number of nodes with $k_{AA}$ and $k_{BB}$ incoming
intra-links to $A$ and $B$ respectively. Finally, we also have
$N_{BA}(k_{BA},t)$ and $N_{AB}(k_{AB},t)$ to denote the average
number of nodes in $A$ and $B$ with $k_{BA}$ and $k_{AB}$ incoming
cross-links.

To keep this paper less cumbersome we will only analyse the time
evolution of network $A$ and apply our results to network $B$. In
addition to this, we only need to give the time evolution of
$N_{A}(k_{AA},k_{BA},t)$, defined as the joint distribution of
intra-links and cross-links. Using this distribution we can find
all other distributions that are mentioned earlier. The time
evolution of $N_{A}(k_{AA},k_{BA},t)$ can be described by a rate
equation
\begin{eqnarray}
\label{Na} \frac{dN_{A}(k_{AA},k_{BA},t)}{dt}&=&\frac{1}{M_{A}(t)}
\{p_{AA}m_{A}[(k_{AA}-1+k_{BA}+a)N_{A}(k_{AA}-1,k_{BA},t)\nonumber\\
&&-(k_{AA}+k_{BA}+a)N_{A}(k_{AA},k_{BA},t)]\nonumber\\
&&+p_{BA}m_{B}[(k_{AA}+k_{BA}-1+a)N_{A}(k_{AA},k_{BA}-1,t)\nonumber\\
&&-(k_{AA}+k_{BA}+a)N_{A}(k_{AA},k_{BA},t)]\}+
\delta_{k_{AA}0}\delta_{k_{BA}0}.
\end{eqnarray}
\noindent The form of the Eq.~(\ref{Na}) seems very similar to the
one used in \cite{geoff}. In that model the rate of creating links
depends on the out-degree of the issuing nodes and the in-degree
of the target nodes. Here we are concerned with two different
types of in-degrees namely intra- and cross-links of the nodes.

On the right hand side of Eq.~(\ref{Na}) the terms in first square
brackets represent the increase in the number of nodes with
$k_{A}$ links when a node with $k_{AA}-1$ intra-links acquires a
new intra-link and if the node already has $k_{A}$ links this
leads to reduction in the number. Similarly, for the second square
brackets where the number of nodes with $k_{A}$ links changes due
to the incoming cross-links. The final term accounts for the
continuous addition of new nodes with no incoming links, each new
node could be thought of as the new publication in a particular
research discipline. The normalization factor $M_{A}(t)$ sum of
all degrees is defined as

\begin{eqnarray}
\label{mat}
M_{A}(t)=\sum_{k_{A}=0}^{\infty}A_{k_{A}}N_{A}(k_{A},t).\end{eqnarray}

\noindent We limit ourself to the case of preferential linear
attachment rate\cite{organization}
\begin{eqnarray}\label{A-rate} A_{k_{A}}=a+k_{A}, \end{eqnarray}
\noindent shifted by $a>0$,  the initial attractiveness
\cite{exactsol} of nodes in $A$, which ensures that there is a
nonzero probability of any node acquiring a link. The nature of
$A_{k_{A}}$ lets one to obtain, as $t \to \infty$
\begin{eqnarray}\label{Mat}
M_{A}(t)=(a+<m_{A}>)t,
\end{eqnarray} where $<m_{A}>=p_{AA}m_{A}+p_{BA}m_{B}$ is the
average total in-degree in Network $A$. Eq.~(\ref{Mat}) implying
that $M_{A}(t)$ is linear in time. Similarly, it is easy to show
that $N_{A}(k_{AA},k_{BA},t)=n_{A}(k_{AA},k_{BA})t$ is also linear
function of time. We use these relations in Eq.~(\ref{Na}) to
obtain the time independent recurrence relation

\begin{eqnarray}\label{arec}
[a+<m_{A}>+<m_{A}>(k_{AA}+k_{BA}+a)]n_{A}(k_{AA},k_{BA})\nonumber\\
=p_{AA}m_{A}(k_{AA}+k_{BA}+a-1)n_{A}(k_{AA}-1,k_{BA})\nonumber\\
+p_{BA}m_{B}(k_{AA}+k_{BA}+a-1)n_{A}(k_{AA},k_{BA}-1)\nonumber\\
+(a+<m_{A}>)\delta_{k_{AA}0}\delta_{k_{BA}0}.
\end{eqnarray}
The expression in Eq.~(\ref{arec}) does not simplify however, it
lets us to obtain the total in-degree distribution
\begin{eqnarray}
\label{tot-in-deg}
N_{A}(k_{A},t)=\sum_{k_{AA}=0}^{k_{A}}N_{A}(k_{AA},k_{A}-k_{AA},t).
\end{eqnarray}
Writing $ N_{A}(k_{A},t)=n_{A}(k_{A})t$ and since
$k_{A}=k_{AA}+k_{BA}$ then $ n_{A}(k_{A})$ satisfies
\begin{eqnarray}\label{narec}
[a+<m_{A}>+<m_{A}>(k_{A}+a)]n_{A}(k_{A})=<m_{A}>(k_{A}+a-1)n_{A}(k_{A}-1)
\nonumber\\ +(a+<m_{A}>)\delta_{k_{A}0}.
\end{eqnarray}

Solving Eq.~(\ref{narec}) for $k_{A}>0$ yields,
\begin{eqnarray}\label{nagamma}
n_{A}(k_{A})=\frac{\Gamma(k_{A}+a)\Gamma(a+2+\frac{a}{<m_{A}>})}
             {\Gamma(k_{A}+a+2+\frac{a}{<m_{A}>})\Gamma(a)}n_{A}(0),
\end{eqnarray}
with
\begin{eqnarray}\label{an0}
n_{A}(0)=\left(1+\frac{<m_{A}>a}{<m_{A}>+a}\right)^{-1}.
\end{eqnarray}

\noindent As $k_{A}\to \infty$ Eq.~(\ref{nagamma}) gives the
asymptotic behaviour of the
 total in-degree distribution in $A$
\begin{eqnarray}\label{ainfty}
n_{A}(k_{A})\sim k_{A}^{-\gamma_{A}},
\end{eqnarray}
\noindent which is a power-law form with an exponent
$\gamma_{A}=2+a/<m_{A}>$ that only depends on the average total
in-degree and the initial attractiveness of the nodes. Similarly,
we can write the total in-degree distribution in network $B$ for
the asymptotic limit of $k_{B}$ as
\begin{eqnarray}\label{binfty}
n_{B}(k_{B})\sim k_{B}^{-\gamma_{B}} \quad \text{with}\quad
\gamma_{B}=2+b/<m_{B}>.
\end{eqnarray}
\noindent Again, the exponent depends upon the initial
attractiveness $b$ of nodes and the average total incoming links
$<m_{B}>=p_{BB}m_{B}+p_{AB}m_{A}$.

We now move on to analyse $N_{AA}(k_{AA},t)$, the distribution of
the average number of nodes with $k_{AA}$ intra-links in network
$A$. In citation network one can think of  these links being
issued from the same subject class as the receiving nodes and in
the case of human sexual contact network, they represent the
homosexual interactions. Since
\begin{eqnarray}
\label{Naa}
N_{AA}(k_{AA},t)=\sum_{k_{BA}=0}^{\infty}N_{A}(k_{AA},k_{BA},t),
\end{eqnarray}
which can also be written as $ N_{AA}(k_{AA},t)=n_{AA}(k_{AA})t$,
a linear function of time. Then summing Eq.~(\ref{arec}) over all
possible values of $k_{BA}$
\begin{eqnarray}\label{naa}
n_{A}(k_{AA})=\sum_{k_{BA}=0}^{\infty}n_{A}(k_{AA},k_{BA}),
\end{eqnarray}
we get
\begin{eqnarray}\label{aarec}
[a+<m_{A}>+<m_{A}>(k_{AA}+a)]n_{AA}(k_{AA})+<m_{A}>
\sum_{k_{BA}=0}^{\infty}k_{BA}n_{A}(k_{AA},k_{BA})\nonumber\\
=p_{AA}m_{A}(k_{AA}+a-1)n_{AA}(k_{AA}-1)+p_{AA}m_{A}
\sum_{k_{BA}=0}^{\infty}k_{BA}n_{A}(k_{AA}-1,k_{BA})\nonumber\\
+p_{BA}m_{B}(k_{AA}+a)n_{AA}(k_{AA})+p_{BA}m_{B}
\sum_{k_{BA}=0}^{\infty}(k_{BA}-1)n_{A}(k_{AA},k_{BA}-1)\nonumber\\
+(a+<m_{A}>)\delta_{k_{AA}0}.
\end{eqnarray}
\noindent For  large $k_{AA}$ Eq.~(\ref{aarec}) reduces to
\begin{eqnarray}\label{re-aarec}
[a+<m_{A}>+p_{AA}m_{A}(k_{AA}+a)]n_{AA}(k_{AA})=\nonumber\\
p_{AA}m_{A}(k_{AA}+a-1)n_{AA}(k_{AA}-1)+(a+<m_{A}>)\delta_{k_{AA}0}.
\end{eqnarray}
\noindent Iterating former relation for $k_{AA}>0$ yields

\begin{eqnarray}\label{aagamma}
n_{AA}(k_{AA})&=&\frac{\Gamma(k_{AA}+a)\Gamma(a+2+
\frac{p_{BA}m_{B}+a}{p_{AA}m_{A}})}
             {\Gamma(k_{AA}+a+2+\frac{p_{BA}m_{B}+a}
             {p_{AA}m_{A}})\Gamma(a)}n_{AA}(0),
\end{eqnarray}
\noindent where
\begin{eqnarray}\label{aan0}
n_{AA}(0)=(1+\frac{p_{AA}m_{A}a}{a+<m_{A}>})^{-1}.
\end{eqnarray}
In the asymptotic limit as $k_{AA} \to \infty$ Eq.~(\ref{aagamma})
has a power-law form
\begin{eqnarray}\label{aainfty}
n_{AA}(k_{AA})\sim k_{AA}^{-\gamma_{AA}}, \quad \text{with
exponent}\quad \gamma_{AA}=2+\frac{p_{BA}m_{B}+a}{p_{AA}m_{A}}
\end{eqnarray}
that depends upon both $p_{AA}$ and the coupling parameter
$p_{BA}$.

Similarly, the time independent recurrence relation for
$N_{BB}(k_{BB},t)$ has the same form as Eq.~(\ref{aarec}) with the
only difference being the parameters. Therefore we will simply
give the power-law distribution
\begin{eqnarray}\label{bbinfty}
n_{BB}(k_{BB})\sim k_{BB}^{-\gamma_{BB}}\quad \text{with} \quad
\gamma_{BB}=2+\frac{p_{AB}m_{A}+b}{p_{BB}m_{B}},
\end{eqnarray}
\noindent where the other coupling parameter $p_{AB}$ is revealed
in the exponent.

Finally, the distribution of average number of nodes with incoming
cross-links $N_{BA}(k_{BA},t)$ in $A$ can be found by summing over
$N_{A}(k_{AA},k_{BA},t)$ for all its intra-links
\begin{eqnarray}
\label{cros-in-deg}
N_{BA}(k_{BA},t)=\sum_{k_{AA}=0}^{\infty}N_{A}(k_{AA},k_{BA},t).
\end{eqnarray}
\noindent As before $N_{BA}(k_{BA},t)=n_{BA}(k_{BA})t$ is also
linear in time. When the cross links $k_{BA}$ are large enough,
then from Eq.~(\ref{arec}) we obtain

\begin{eqnarray}\label{bagamma}
n_{BA}(k_{BA})&=&\frac{\Gamma(k_{BA}+a)\Gamma(a+2+
\frac{p_{AA}m_{A}+a}{p_{BA}m_{B}})}
             {\Gamma(k_{BA}+a+2+\frac{p_{AA}m_{A}+a}
             {p_{BA}m_{B}})\Gamma(a)}n_{BA}(0),
\end{eqnarray}
\noindent where
\begin{eqnarray}\label{ban0}
n_{BA}(0)=(1+\frac{p_{BA}m_{B}a}{<m_{A}>+a})^{-1}.
\end{eqnarray}
\noindent In the asymptotic limit as $k_{AB}\to \infty $ the
distribution
\begin{eqnarray}\label{bainfty}
n_{BA}(k_{BA})\sim k_{BA}^{-\gamma_{BA}}, \quad
\gamma_{BA}=2+\frac{p_{AA}m_{A}+a}{p_{BA}m_{B}}
\end{eqnarray} has a power-law form and similarly
for the network $B$ as $k_{AB}\to \infty $
\begin{eqnarray}\label{abinfty}
n_{AB}(k_{AB})\sim k_{AB}^{-\gamma_{AB}} \quad \text{with} \quad
\gamma_{AB}=2+\frac{p_{BB}m_{B}+b}{p_{AB}m_{A}}.
\end{eqnarray}
\noindent Unlike the case in intra-links, here the exponents are
inversely proportional to the coupled parameters $p_{BA}$ and
$p_{AB}$ respectively.
\section{Discussion and conclusions}
\label{sec:conclusions}

For the sake of simplicity, we set the number of outgoing links of
the new nodes in either networks to be the same, i.e.
$m_{A}=m_{B}=m$. Furthermore taking the rate of cross linking to
be $p_{AB}=p_{BA}=q$ and the rate of intra linking
$p_{AA}=p_{BB}=p$, consequently we have $p=1-q$, and $q$ as the
coupling parameter.

In the weak coupling case, the cross linking is negligibly small
i.e. $q\to 0$ then the power-law exponent of the intra-link
distribution is
\begin{eqnarray}\label{week-coupling}
\gamma_{AA}=2+\frac{qm+a}{(1-q)m}\end{eqnarray}  equal to total
link distribution $\gamma_{AA}=2+a/m$. This gives a solution
obtained in\cite{exactsol} and when $a=m$ we recover the exponent
$\gamma=3$, the empirical findings in \cite{citation}.

The case of strong coupling $q \to 1$ is an illustrative example
for citation networks, which results in $\gamma_{AA}=\gamma_{BB}
\to \infty$, both intra-link distributions having exponents
approaching to infinity.

Thus, varying $q$ in $(0,1)$ yields any values of $\gamma_{AA}$
between $(2+a/m)$ and $\infty$. On the contrary, the exponent of
cross-link distribution
\begin{eqnarray}\label{cross-exp}\gamma_{BA}=2+\frac{(1-q)m+a}{qm}
\end{eqnarray} decreases from $\infty$ to
$2+a/m$, as $q$ increases from $0$ to $1$.

Taking $a=m$ gives
\begin{eqnarray}\label{a=m}\gamma_{AA}=2+\frac{1+q}{1-q}
\quad \text{and} \quad \gamma_{BA}=1+\frac{2}{q}.\end{eqnarray}
Supposing $0<q<0.5$, which seems reasonable for consideration of
citation networks, we find that $3<\gamma_{AA}<5$ and
$\gamma_{BA}>5$. The former result coincides with the distribution
of connectivities for the electric power grid of Southern
California \cite{barab,amaral}. Where the system is small and the
local interactions is of importance hence there seems to be some
analogy to the intra-linking process. For the latter, as far as we
are aware there is none empirical studies present in the published
literature.

Now, consider the web of human sexual
contacts\cite{liljeros,guler}. If we let $A$ to represent males
and $B$ females that is $A=M$ and $B=F$ then
\begin{eqnarray}\gamma_{FM}=2+\frac{(1-q)m+a}{qm}\quad \text{and}
\quad \gamma_{MF}=2+\frac{(1-q)m+b}{qm} \end{eqnarray} are the
power-law exponents of the degree distributions of the sexes.
Where $a$ and $b$ denote the male and female attractiveness
respectively and usually $a<b$ is considered \cite{guler}. By
setting $a/m=1.31$, $b/m=1.54$ and $q \to 1$ that is, cross links
are predominant then as in \cite{guler} we obtain
$\gamma_{FM}=1+\alpha_{FM}=3.31$ for males and
$\gamma_{MF}=1+\alpha_{MF}=3.54$ for females. The exponents
$\alpha_{FM}=2.31$ and $\alpha_{MF}=2.54$ have been observed for
the cumulative distributions in empirical study \cite{liljeros}.

The model we studied here seems to have the flexibility to
represent variety of complex systems.
\acknowledgments We would like to thank the China Scholarship
Council, EPSRC for their financial support and Geoff Rodgers for
useful discussions.


\end{document}